\documentclass[journal]{IEEEtran}
\usepackage{cite}
\usepackage{lineno}
\usepackage[colorlinks,citecolor=cvprblue]{hyperref}
\usepackage{color}
\usepackage{array}
\usepackage{subfigure}
\usepackage{url}
\usepackage{amsmath,amssymb,amsfonts}
\usepackage{algorithmic}
\usepackage{graphicx}
\usepackage{textcomp}
\usepackage{booktabs}
\usepackage{multirow}
\usepackage{orcidlink}
\usepackage{colortbl}
\usepackage{pifont}
\usepackage[capitalize]{cleveref}
\crefname{section}{Sec.}{Secs.}
\Crefname{section}{Section}{Sections}
\Crefname{table}{Table}{Tables}
\crefname{table}{Tab.}{Tabs.}
\definecolor{cvprblue}{rgb}{0.21,0.29,0.94}
\ifCLASSINFOpdf
\else
\fi
\begin{document}
%
\title{CoCPF: Coordinate-based Continuous Projection Field for Ill-Posed Inverse Problem in Imaging}
%
%
%
\author{Zixuan~Chen\textsuperscript{\orcidlink{0000-0001-5100-793X}},~
        Lingxiao~Yang,~
        Jian-Huang~Lai\textsuperscript{\orcidlink{0000-0003-3883-2024}},~\IEEEmembership{Senior Member,~IEEE},~
        and~Xiaohua~Xie\textsuperscript{\orcidlink{0000-0002-0310-4679}}
\thanks{Manuscript received XXX XX, XXXX; revised XXX XX, XXXX. This work was supported in part by the National Natural Science Foundation of China under Grant 62072482.
}
\thanks{(Corresponding author: Xiaohua Xie.)}
\thanks{The authors are with the School of Computer Science and Engineering, Sun Yat-sen University, Guangzhou 510006, China; and with the Guangdong Province Key Laboratory of Information Security Technology, Guangzhou 510006, China; and also with the Key Laboratory of Machine Intelligence and Advanced Computing, Ministry of Education, Guangzhou 510006, China. (e-mail: chenzx3@mail2.sysu.edu.cn; \{yanglx9, stsljh, xiexiaoh6\}@mail.sysu.edu.cn)}
}

%
%

\markboth{IEEE TRANSACTIONS ON IMAGE PROCESSING,~Vol.~x, No.~x, May~2020}%
{Shell \MakeLowercase{\textit{et al.}}: Bare Demo of IEEEtran.cls for IEEE Journals}
%



\newcommand{\bt}[1]{\textbf{#1}}
\newcommand{\ut}[1]{\underline{#1}}
\newcommand{\df}[1]{\scriptsize{\textcolor{Gray}{\textbf{#1}}}}
\newcommand{\xt}[1]{\textit{#1}}
\newcommand{\ua}{$\uparrow$}
\newcommand{\da}{$\downarrow$}
\newcommand{\ck}{\ding{52}}
\newcommand{\tr}[1]{\textcolor{red}{#1}}
\newcommand{\ty}[1]{\tiny #1}
\newcommand{\etal}{\xt{et al.\ }}
\definecolor{Gray}{rgb}{0.35, 0.35, 0.35}
\definecolor{MyGray}{rgb}{0.9, 0.9, 0.9}
\definecolor{DarkGray}{rgb}{0.8, 0.8, 0.8}
\newcommand{\cc}{\cellcolor{MyGray}}
\newcommand{\dc}{\cellcolor{DarkGray}}
\newcommand{\link}[1]{\href{#1}{#1}}
\maketitle

\begin{abstract}
Sparse-view computed tomography (SVCT) reconstruction aims to acquire CT images based on sparsely-sampled measurements.
It allows the subjects exposed to less ionizing radiation, reducing the lifetime risk of developing cancers.
Recent researches employ implicit neural representation (INR) techniques to reconstruct CT images from a single SV sinogram. 
However, due to ill-posedness, these INR-based methods may leave considerable ``holes'' (\xt{i.e.,} unmodeled spaces) in their fields, leading to sub-optimal results.
In this paper, we propose the Coordinate-based Continuous Projection Field (CoCPF), which aims to build hole-free representation fields for SVCT reconstruction, achieving better reconstruction quality.
Specifically, to fill the holes, \xt{CoCPF} first employs the \xt{stripe-based volume sampling} module to broaden the sampling regions of \xt{Radon transformation} from rays (1D space) to stripes (2D space), which can well cover the internal regions between SV projections.
Then, by feeding the sampling regions into the proposed differentiable rendering modules, the holes can be jointly optimized during training, reducing the ill-posed levels.
As a result, \xt{CoCPF} can accurately estimate the internal measurements between SV projections (\xt{i.e.,} DV sinograms), producing high-quality CT images after re-projection.
Extensive experiments on simulated and real projection datasets demonstrate that \xt{CoCPF} outperforms state-of-the-art methods for 2D and 3D SVCT reconstructions under various projection numbers and geometries, yielding fine-grained details and fewer artifacts. Our code will be publicly available.

\end{abstract}

\begin{IEEEkeywords}
  Imaging Inverse Problem, Sparse-View Computed Tomography Reconstruction, Implicit Neural Representation, Self-supervised Learning, Radon Transformation.
\end{IEEEkeywords}

%
\IEEEpeerreviewmaketitle

\section{Introduction}
%
%
%
%

\IEEEPARstart{C}{omputed} tomography (CT), a non-invasive and non-destructive tool for observing the internal structure of scanned objects, is widely-used to obtain visual information for assisting clinical diagnosis.
In CT imaging, the photons emitted from an X-ray-generating source can bring information about attenuation properties to the rotating detectors around a subject.
The detected projection profiles at each angle are stored in the sinograms.
To reconstruct a density field related to the scanned object from the sinograms, one commonly-used approach is the filtered back-projection (FBP) \cite{FBP}, an analytical reconstruction technique for the inversion.
However, the acquisition of high-quality CT images requires densely-sampled measurements, so subjects need to be exposed to considerable ionizing radiation for a long time, increasing the lifetime risk of developing cancers \cite{CT}.
To reduce ionizing radiation brought from CT imaging, one strategy is to reconstruct CT images from sparsely-sampled projection views, \textit{i.e.,} Sparse-View Computed Tomography (SVCT) reconstruction, which is a pressing demand for public healthcare and receiving widespread concerns in medical domains.

The acquisition of SV sinogram $y$ can be formulated as:
\begin{equation}
  y = A\mathbf{x}+e,
\end{equation}
where $A\in\mathbb{R}^{m\times n}$ ($m\ll n$) is a sparse sampler, and $\mathbf{x}\in\mathbb{R}^{n}$ and $y\in\mathbb{R}^{m}$ are full- and sparse-view measurements, respectively. $e\in\mathbb{R}^{m}$ denotes the additive white Gaussian (AWGN) noise.
Since the number of linearly-independent measurements (rank of $A$) is less than the number of unknowns (length of $\mathbf{x}$), this inverse problem has infinitely-many solutions (\textit{i.e.,} ill-posed problem), and directly using the re-projection techniques \cite{FBP,ART} may lead to severe artifacts.

Early studies \cite{art4svct,fista,FISTA-TV} formulate the ill-posed inverse problem as the regularized optimizations.
With the advent of deep learning, a series of methods employ the Convolutional Neural Networks (CNNs) to solve that problem, which can be split into two groups: \bt{i)} \xt{CNN-based denoisers} \cite{FBPConv,TFUnet,metainv,ddnet,DPCCT,FISTA-Net,DSAL}, treating the artifacts as noise and using the CNN-based denoising models for removal; and
\bt{ii)} \xt{View-synthesis CNNs} \cite{view-syn1,DSS,DRONE,freeseed,GMSD}, yielding the dense-view (DV) sinograms $\hat{\mathbf{x}}\in\mathbb{R}^n$ by the CNN-based upsamplers $\mathcal{M}_{\varsigma}$ and acquire the CT images $\hat{x}$ by:
\begin{equation}
    \hat{x}=\mathcal{R}(\mathcal{M}_{\varsigma}(y)), 
    \label{eq:view-syn-cnns}
\end{equation}
where $\mathcal{R}$ denotes the re-projection techniques.

\begin{figure}[!t]
  \centering
  \includegraphics[width=0.48\textwidth]{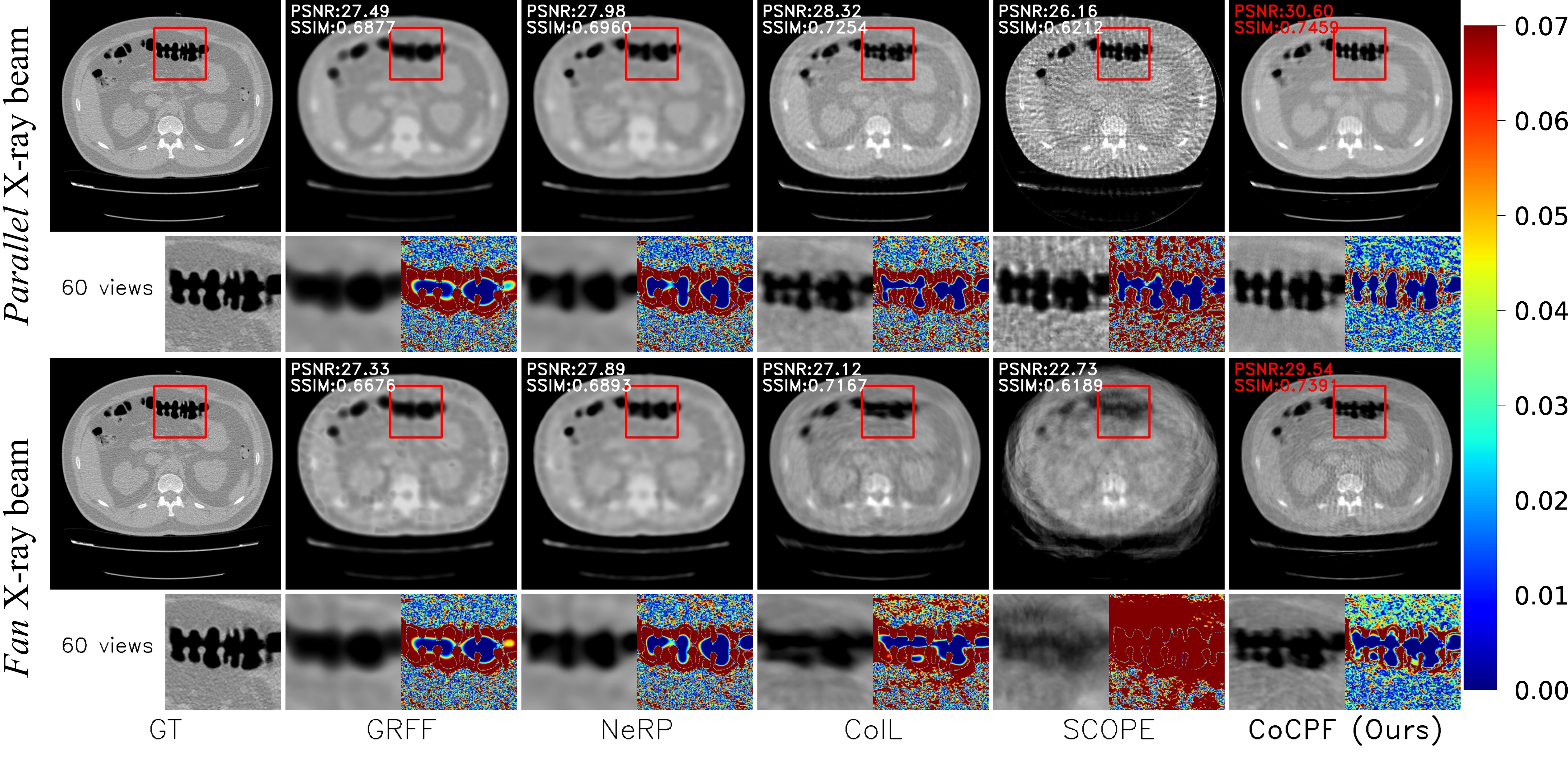}
  \caption{
    Visual examples against state-of-the-art INRs: GRFF \cite{GRFF}, NeRP \cite{NeRP}, CoIL \cite{CoIL} and SCOPE \cite{scope} for \bt{2D} SVCT reconstruction on COVID-19 \cite{covid19} dataset.
    Each subfigure (bottom left) highlights the anatomic structures zoomed in the boxes, while the heatmap (bottom right) shows the difference related to the GT.
    \textcolor{red}{Red} text indicates the highest score.
  }
  \label{fig:examples}
\end{figure}

Recent researches \cite{GRFF,NeRP,CoIL,scope} utilize the implicit neural representation (INR) techniques to build the mappings between the positional coordinates and the corresponding pixel values, which can be trained on a single SV sinogram to acquire the corresponding CT image without extra information, significantly saving training costs.
However, existing INR-based methods only build the mappings between coordinates and SV sinograms based on \xt{Radon transformation}, which cannot reduce the ill-posed levels brought by sparse samplers.
After optimization, their fields may form some ``holes'', \xt{i.e., the regions that have never been modeled.}
As shown in \cref{fig:examples}, these holes may produce blurry contents (GRFF \cite{GRFF} and NeRP \cite{NeRP}) and severe artifacts (CoIL \cite{CoIL} and SCOPE \cite{scope}).

To address the above issues, we present \bt{Co}ordinate-based \bt{C}ontinuous \bt{P}rojection \bt{F}ield (CoCPF), a hole-free representation field for SVCT reconstruction.
Specifically, to fill the holes, \xt{CoCPF} first employs the \xt{stripe-based volume sampling} module to broaden the sampling regions of \xt{Radon transformation} from rays (1D space) to stripes (2D space), which can well cover the internal regions between SV projections by the spatial overlaps.
Then, by feeding the sampling regions into the proposed differentiable rendering modules: xt{piecewise-consistent volume rendering} and \xt{stripe-based hierarchical volume rendering} modules, the holes can be jointly optimized during training, reducing the ill-posed levels.
As a result, \xt{CoCPF} can accurately estimate the representation within the internal regions between adjacent SV projections (\xt{i.e.,} DV sinograms), producing high-quality CT images after re-projection.
Compared with existing INR-based methods, our \xt{CoCPF} achieves higher reconstruction quality, including better visual verisimilitude and fewer artifacts (see \cref{fig:examples}).
Comprehensive experiments on two simulated datasets: COVID-19 \cite{covid19} and KiTS19 \cite{kits19}, and a real projection dataset: AAPM-LDCT-PD \cite{aapm} demonstrate that \xt{CoCPF} outperforms state-of-the-art INR- and CNN-based methods for 2D and 3D SVCT reconstructions, achieving consistently preferable performance under various projection numbers and geometries. 

The main contributions can be summarized as follows:


\begin{itemize}
  \item We reveal existing INR-based methods suffer from hole-forming issues due to the ill-posedness of SVCT reconstruction, and present \xt{CoCPF} to build the hole-free representation fields for better reconstruction quality.
  
  \item We propose \xt{stripe-based volume sampling}, \xt{piecewise-consistent volume rendering}, and \xt{stripe-based hierarchical rendering} modules to address the hole-forming issues, reducing the ill-posed levels brought by the sparsely-sampled measurements for SVCT reconstruction.
  
  \item We conduct a series of experiments to demonstrate the effectiveness and correctness of our \xt{CoCPF}, which surpasses the state-of-the-art methods on simulated and real projection datasets, yielding consistently superior performance under various projection numbers and geometries. 
  
\end{itemize}

The remainder of this paper is structured as follows. 
Related works and preliminaries are reviewed in \Cref{sec:rw,sec:pre}, respectively. \Cref{sec:method} presents the analysis, motivation, and details of our approach. \Cref{sec:exp} demonstrates the experimental results and ablation study on three public datasets, and conclusions are drawn in \Cref{sec:con}.

\section{Related Works}\label{sec:rw}
In this section, we first review the implicit neural representation (INR) techniques and then introduce the advances in sparse-view computed tomography (SVCT) reconstruction.

\begin{figure*}[!t]
  \centering
  \includegraphics[width=\textwidth]{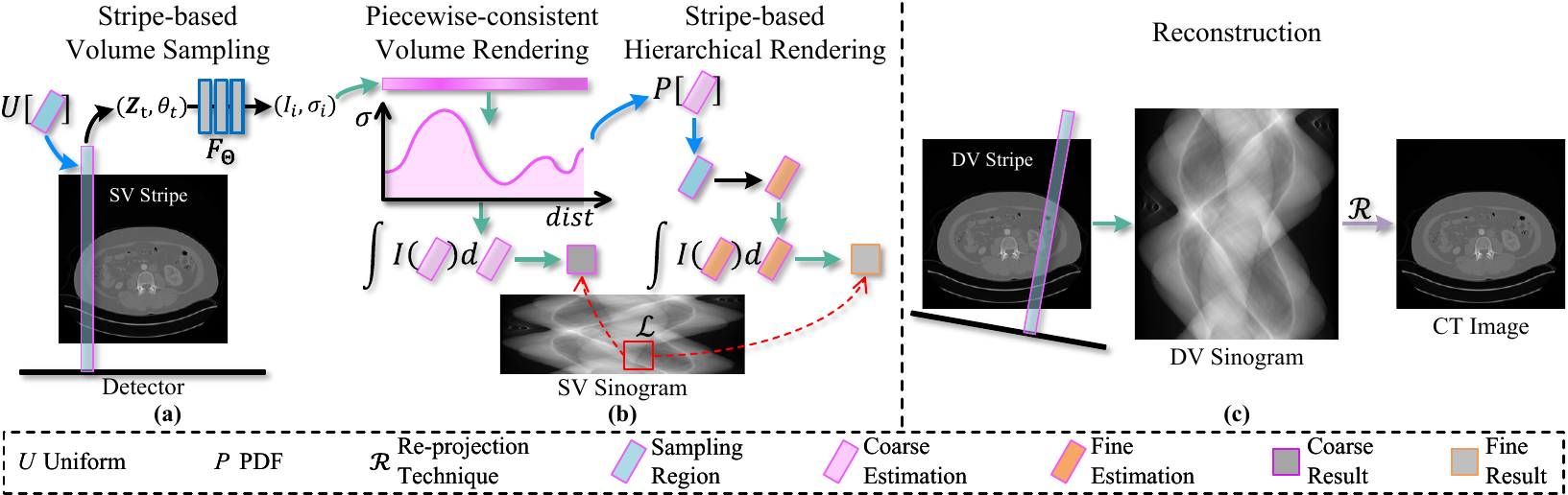}
  \caption{
    The overall architecture of the proposed \xt{CoCPF}.
    In training stage, given a projection angle $\theta_t$ from sparse angle partitions, \bt{(a)} \xt{CoCPF} uniformly samples the points within a stripe $\mathcal{S}_{k}(\varpi, \rho, \theta_{t})$ (blue stripe).
    By feeding the stripe coordinate $\mathbf{z}_{i}$ and $\theta_{t}$ of each sampling point into the MLP $F_{\Theta}$, \xt{CoCPF} produces the corresponding light intensity $I_i\in\mathbb{R}$ and attenuation coefficient $\sigma_i\in\mathbb{R}$.
    \bt{(b)} As discussed in \cref{eq:simple}, the distribution of stripe is only related to the distance $\nu$, \xt{CoCPF} predicts a coarse result (purple box) by employing the volume integral \cref{eq:M_volume} on the coarse estimations (pink stripe) of that stripe.
    Then, \xt{CoCPF} resamples the points under the PDF of coarse estimation to acquire the fine one (orange stripe), and the fine result (orange box) can be generated by the former process.
    We use the fine result as the final output.
    Since the rendering functions are differentiable, \xt{CoCPF} can optimize the MLPs by minimizing the loss \cref{eq:loss} between the dual outputs and sinogram pixels.
    \bt{(c)} After optimization, CT images can be acquired by applying FBP \cite{FBP} on the synthesized DV sinograms as \cref{eq:view-syn-cnns}.
  }
  \label{fig:overall}
  \end{figure*}

\subsection{Implicit Neural Representation}
Learning a continuous function from discrete samples is a long-standing research problem in computer vision for numerous tasks. 
One recent trend is to build a coordinate-based representation field from sparsely-sampled data using implicit neural representation (INR) techniques.
This coordinate-based representation field can be regarded as a continuous function, which can yield identically distributed samples in a continuous domain.
Based on the image continuity priors, INR techniques have yielded impressive performance in various vision problems.
One compelling strategy is neural radiance fields (NeRF) \cite{NeRF}, which yielded impressive view synthesis based on conventional photos for 3D reconstruction.
Using the standard volumetric rendering \cite{rendering} based on the Lambert-Beer Law, NeRF \cite{NeRF} can model a radiance field to render novel views with high visual quality.
Follow-up works adapted NeRF \cite{NeRF} to the various domains, such as generative modeling \cite{graf}, editing \cite{nerf-editing}, and anti-aliasing \cite{mip-nerf}.
Recently, some researchers have attempted to adapt INR to the medical domain. 
Corona-Figueroa \etal \cite{mednerf} employ \cite{graf} to synthesize novel-view X-rays with training on multi-view X-rays, while Chen \etal \cite{chen2023cunerf} propose the cube-based modeling strategy to upsample medical images at arbitrary scales in a continuous domain.
More details can be seen in the recent survey \cite{INR4MedicalSurvey}.

\subsection{Sparse-view Computed Tomography Reconstruction}
Sparse-view computed tomography (SVCT) reconstruction is a challenging ill-posed problem in medical image reconstruction, which aims to reconstruct a density field of the scanned subject from the sparsely-sampled projections (\xt{i.e.,} SV sinogram).
Initially, Filtered Back-Projection (FBP) \cite{FBP} and Algebraic Reconstruction Technique (ART) \cite{ART} are employed in early research.
To eliminate the artifacts caused by ill-posedness, early studies \cite{art4svct,fista,FISTA-TV} model this inverse problem as the regularized optimization.
Feng \etal \cite{art4svct} propose to optimize the least-square and total variance (TV) from the discrete gradient operator \cite{tv}, while \cite{fista,FISTA-TV} solve the nonsmooth optimization problems by the \textit{proximal operator} \cite{prox}.
Subsequently, the reconstruction quality has been significantly improved by the Convolutional Neural Networks (CNNs), which can be split into two groups: \textit{i)} CNN-based denoisers \cite{FBPConv,TFUnet,metainv,ddnet,DPCCT,FISTA-Net,DSAL}, aiming to learn the denoising patterns from the mappings between sparse- and full-view CT pairs.
Specifically, \cite{FBPConv,TFUnet,metainv} employ U-Net \cite{U-net} to fuse hierarchical features, Zhang \etal \cite{ddnet}, Fu \etal \cite{DPCCT} merge the features from different dense blocks \cite{densenet}, while Xiang \etal \cite{FISTA-Net} and Lahiri \etal \cite{DSAL} iteratively cascade the convolution blocks during optimization; and \textit{ii)} View-synthesis CNNs \cite{view-syn1,DSS,DRONE,freeseed,GMSD}, yielding DV sinograms from the SV ones for CT acquisition.
For sinogram super-resolution, \cite{view-syn1,DSS} utilize the linear interpolations, \cite{DRONE,freeseed} employ U-Net-based SR models, and \cite{GMSD} reform the score-based diffusion model \cite{SDE}. 
Recent studies apply the INR techniques on a single SV sinogram to reconstruct CT images.
Tancik \etal \cite{GRFF} propose an efficient Fourier feature mapping to encode the coordinates, while Shen \etal \cite{NeRP} utilize the prior images for further improvements.
To acquire DV sinograms, Sun \etal \cite{CoIL} build the point-to-point mapping between the coordinates and the sinogram pixels, and Wu \etal \cite{scope} simulate the X-rays to build this mapping.

\section{Preliminary}\label{sec:pre}
Implicit Neural Representations (INRs) aim to model a continuous function from discrete samples by the mappings between the coordinates and pixel values, which can be used to reconstruct CT images based on a single SV sinogram.
These INR-based methods can be split into two groups:

\noindent\bt{Direct-reconstruction INRs} \cite{GRFF,NeRP}\bt{.}
Given an SV sinogram $y$, they first simulate the CT density field $\hat{x}$ by the coordinates and then use the differentiable forward-projection $\tilde{A}$ to optimize the parameters $\Theta$ as:
\begin{equation}
    \Theta^*=\mathop{\arg\min}\limits_{\Theta}\|\tilde{A}\hat{\mathbf{x}}-y\|^2_2.
    \label{eq:DRINR}
\end{equation}

\noindent\bt{View-synthesis INRs} \cite{CoIL,scope}\bt{.}
They aim to feed the coordinates into $F_{\Theta}$ to yield the corresponding projection views, which can be optimized by the synthesized SV sinograms $\hat{y}$:
\begin{equation}
  \Theta^*=\mathop{\arg\min}\limits_{\Theta}\|\hat{y}-y\|^2_2.
  \label{eq:VSINR}
\end{equation}
After optimization, \xt{view-synthesis INRs} can acquire CT images by applying the re-projection techniques $\mathcal{R}$ on the synthesized DV sinograms like \cref{eq:view-syn-cnns}.
\xt{Note}, they can be integrated into various image reconstruction techniques, including the re-projection and denoising methods, for further improvements.

\section{Method}\label{sec:method}
In this section, we first analyze the reasons why existing INR-based methods suffer from hole-forming issues.
To address this problem, we propose \bt{Co}ordinate-based \bt{C}ontinuous \bt{P}rojection \bt{F}ield (CoCPF), comprising \xt{stripe-based volume sampling}, \xt{piecewise-consistent volume rendering} and \xt{stripe-based hierarchical rendering} modules.
After optimization, the holes can be joint-optimized by the spatial constraints.
As a result, \xt{CoCPF} can reduce the ill-posed levels, acquiring high-quality CT images after re-projection (see results in \cref{fig:examples}).

The overall architecture of \xt{CoCPF} is depicted in \cref{fig:overall}.
Visually, \xt{CoCPF} first builds a coordinate-based representation field and then employs \xt{stripe-based volume sampling} module to sample $N$ points within the stripe.
Subsequently, \xt{CoCPF} utilizes \xt{piecewise-consistent volume rendering} and \xt{stripe-based hierarchical rendering} modules to obtain the coarse and fine results, respectively.
Since the proposed rendering modules are differentiable, the MLP can be optimized by \cref{eq:loss}.
After optimization, we can acquire the CT images by applying FBP \cite{FBP} on the predicted sinograms like \cref{eq:view-syn-cnns}.

\subsection{Analysis \& Motivation} \label{sec:analysis}
\begin{figure}[!t]
  \centering
  \includegraphics[width=0.48\textwidth]{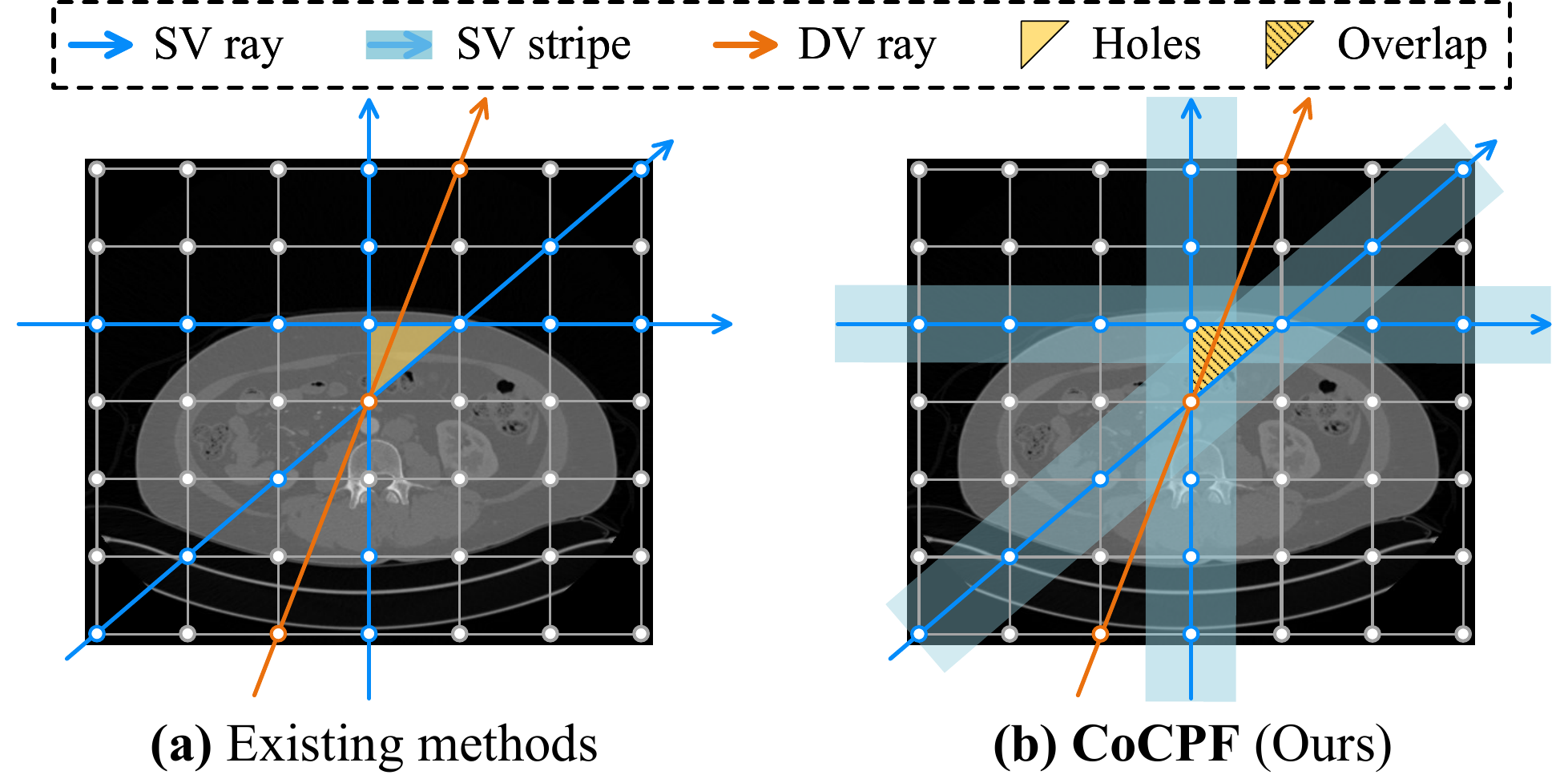}
  \caption{
    \bt{(a)} Existing INR-based methods build the mappings between the positional coordinates and the sinogram pixels based on the \xt{Radon transformation}.
    However, they struggle to reduce the ill-posed levels brought by the sparse sampler, thereby forming the holes (\xt{i.e.,} unmodeled spaces) in the fields, leading to blurry results and severe artifacts.
    \bt{(b)} \xt{CoCPF} broadens the sampling regions from rays (1D space) to stripes (2D space), thereby the holes can be jointly optimized in training, reducing the ill-posed levels for SVCT reconstruction.
  }
  \label{fig:analysis}
\end{figure}

Most INR-based methods directly employ \xt{Radon transformation} to model the mappings between the CT density field coordinates and sinogram pixels.
However, we found that these methods may leave some ``holes'', (\xt{i.e.,} unmodeled spaces) within the representation fields they built.
To explain this issue, we provide an intuitive example in \cref{fig:analysis} \bt{(a)}.
Visually, simulating the rays emitted to the detector can only model the points along the rays, while the internal regions between adjacent rays have never been modeled during optimization, leading to sub-optimal results.
Specifically, as optimized by \cref{eq:DRINR}, the \xt{direct-reconstruction} INRs may maintain some ``tolerable errors'' brought by the ill-posedness, losing high-frequency information and thus yielding blurry contents (see GRFF \cite{GRFF} and NeRP \cite{NeRP} in \cref{fig:examples}).
For the \xt{view-synthesis} INRs, yielding a DV sinogram necessitates querying the holes, resulting in bias estimation in the projection domain.
This bias may be magnified in the re-projection, producing severe artifacts (see CoIL \cite{CoIL} and SCOPE \cite{scope} in \cref{fig:examples}).

To reduce the ill-posed levels brought by the sparsely-sampled measurements, it is required to ``fill'' the holes during optimization.
As depicted in \cref{fig:analysis} \bt{(b)}, broadening the sampling regions of \xt{Radon transformation} from rays (1D space) to stripes (2D space) can cover the internal regions by the spatial overlaps, thereby 
As a result, the holes can be jointly optimized in training, which can reduce the ill-posed levels and yield accurate projection views between the adjacent SV projections, yielding fine-grained details and fewer artifacts after re-projection (see the results in \cref{fig:examples}).

\subsection{Stripe-based Volume Sampling}
As introduced in \Cref{sec:analysis}, due to the ill-posedness caused by sparsely-sampled measurements, existing INR-based methods cannot build the hole-free representation fields, leading to blurry results and severe artifacts.
To address this issue, we propose a novel sampling strategy: \xt{stripe-based volume sampling}, which samples stripes (2D space) instead of rays (1D space) to fill the holes by broadening the sampling regions of \xt{Radon transformation}.
Specifically, we first normalize the coordinate from the image space to the field space within the range of $(-1, 1)$.
For example, the positional coordinate $(x_i, y_i, l_i)$ of a 3D CT volume can be transformed into the field coordinate $\mathbf{z}_{i}$ by:
\begin{equation}
\mathbf{z}_{i} =\left(\frac{2x_{i}-W}{W+2P}, \frac{2y_{i}-H}{H+2P}, \frac{2l_i-L}{L+2P}\right),
\end{equation}
where $P$ is a hyperparameter of the padding size to restrict each $\mathbf{z}_i$'s element within the range of $(-1, 1)$.
$W$ and $H$ denote the size of the detector $D\in\mathbb{R}^{W\times H}$, while $L$ represents the length of the ray casting emitted to the detector.

Then, we construct a base stripe region $\mathcal{S}_{k}(\varpi, \rho, 0^\circ)$ emitted to the detector position $\mathcal{D}_{k}$, whose size is $\varpi\times\rho$ and the projection angle is $0^\circ$.
Given a projection angle $\theta_t$, the base stripe can be transformed into $\mathcal{S}_{k}(\varpi, \rho, \theta_t)$ by the rotation matrix $M(\theta_t)$.
Finally, we sample $N$ points as a point set $\{\mathbf{z}_i\}^N_{i=1}$ within the stripe under the uniform distribution $\mathcal{U}$, and each point can be sampled by:
\begin{equation}
\mathbf{z}_{i}\sim\mathcal{U}\left[\mathcal{S}_{k}(\varpi, \rho, \theta_t)\right].
\end{equation}
The group of these $N$ sampling points is employed to approximate the distribution of the stripe space.

\subsection{Piecewise-consistent Volume Rendering}\label{sec:real}
To estimate the projection views, most INR-based methods \cite{GRFF,NeRP,scope} employ the summation formula to deal with the sampling points.
However, the above formula is more likely to fit the synthesis data based on \xt{Radon transformation} but not the real projection views (see the results on real projection data in \Cref{tab:AAPM} and \cref{fig:visual_aapm}). 
Moreover, this formula ignores the geometry of sampling regions, which may lead to overfitting on the proposed stripe-based volume sampling. 
To simulate the physics of computed tomography and adapt the stripe-based geometry, we reform the volume rendering \cite{rendering,NeRF} to construct the spatial constraints within the sampled stripes based on the Beer-Lambert law.
Specifically, for each sampling point $\mathbf{z}_{i}$ with the projection angle $\theta_t$, we first predict the corresponding light intensity $I(\mathbf{z}_{i}, \theta_t)$ and attenuation coefficient $\sigma(\mathbf{z}_{i})$ using a multi-layer perceptron (MLP) $F_{\Theta}$.
The sinogram result $\mathbf{C}(k, \theta_t)$ related to the stripe region $\mathcal{S}_{k}(\varpi, \rho, \theta_t)$ can be formulated as the double integral:
\begin{equation}
\mathbf{C}(k, \theta_t) = \!\!\!\iint\limits_{\mathcal{S}_{k}(\varpi, \rho, \theta_t)}\!\!\!\frac{I(\mathbf{z}, \theta_t)(1-\exp(-\sigma(\mathbf{z})))}{\exp(\!\!\!\!\!\iint\limits_{\mathcal{S}_{k}(x, y, \theta_t)}\!\!\!\!\!\!\sigma(\mathbf{z}))dx'dy'}dxdy.
\label{eq:volume}
\end{equation}
However, using numerical quadrature to approximate the above double integral requires sampling considerable points within the stripe, leading to massive computational costs.

To simplify the above integral, we propose \xt{piecewise-consistent volume rendering} module, which first splits a stripe $\mathcal{S}_{k}(\varpi, \rho, \theta_t)$ into the pieces by the sorted sampling points.
Then, the attenuation coefficients $\sigma$ and light intensity $I$ within each piece can be approximated as the same.
Thus, the distribution of $\mathcal{S}_{k}(\varpi, \rho, \theta_t)$ can be approximated as a function related to the distance $\nu$, which \cref{eq:volume} can be simplified as:
\begin{equation}
\mathbf{C}(k, \theta_t) = \varpi\int_{0}^{\rho}\frac{I(\nu, \theta_t)(1-\exp(-\sigma(\nu)))}{\exp(\varpi\int_{0}^{\nu}\sigma(\nu'))d\nu'}d\nu,
\label{eq:simple}
\end{equation}
where $\varpi$ denotes the width of the stripe.
Given $N$ sampling points, we sort the points by the distance $\nu$.
The integral can be approximated by the numerical quadrature rules:
\begin{equation}
\hat{\mathbf{C}}(k, \theta_t)=\sum^{N}_{i=1}\frac{\varpi(1-\exp(-\sigma(\nu_i)(\nu_{i+1}-\nu_{i})))}{\exp(\varpi\sum_{j=1}^{i}\sigma(\nu_j)(\nu_{j+1}-\nu_{j}))}I(\nu_i, \theta_t).
\label{eq:M_volume}
\end{equation}

\subsection{Stripe-based Hierarchical Rendering}
To refine the results, \xt{CoCPF} allocates sampling points proportionally to their expected distribution within the stripe regions.
Similar to NeRF \cite{NeRF}, the proposed \xt{CoCPF} also simultaneously optimizes two MLPs, \xt{i.e.}, the coarse one $F^{c}_{\Theta}$ and the fine one $F^{f}_{\Theta}$.
\xt{CoCPF} first samples $N^c$ points and obtain the coarse output $\hat{\mathbf{C}}^c(k, \theta_t)$ by \cref{eq:M_volume}, which can be rewritten as $\hat{\mathbf{C}}^c(k, \theta_t)=\sum^{N^c}_{i=1}w_{i}\cdot I(\nu_i, \theta_t)$.
Thus, the PDF related to a stripe $\mathcal{S}_{k}(\varpi, \rho, \theta_t)$ can be estimated by $\{\hat{w}_i\}^{N^c}_{i=1}$, where $\hat{w}_i = w_{i}/\sum^{N^c}_{j=1}w_{j}$.
Because $\hat{w}_i$ is only related to the distance $\nu_i$, \xt{CoCPF} can resample $\{\nu_i\}^{N^f}_{i=1}$ with $N^f$ elements under the distribution of coarse estimation by the inverse transform sampling (ITS).
Then, \xt{CoCPF} uses the hierarchical sampling function $\zeta_{k}(\cdot)$ to sample the coordinates \xt{w.r.t} $\nu$ within the stripe region $\mathcal{R}_{k}(w, l, \theta_t)$:
\begin{equation}
\mathbf{z}^f_i=\zeta_{k}(\xi, \nu, \theta_t),\ \xi \sim \mathcal{U}([-\frac{\varpi}{2}, \frac{\varpi}{2}])
\end{equation}
where $\xi$ is a random offset and $\zeta_{k}(\cdot)$ is employed to convert the sampled variants $(\xi, \nu, \theta_t)$ into the field sapces.
Finally, these two MLPs $F^{c}_{\Theta}$ and $F^{f}_{\Theta}$ are optimized by minimizing the rendering loss \xt{w.r.t} the coarse and fine results as:
\begin{equation}
\mathcal{L}=\sum_{k, \theta_t\in \mathcal{B}}\left[\lambda\|g.t. - \hat{\mathbf{C}}^c(k, \theta_t)\|^{2}_{2}+\|g.t. - \hat{\mathbf{C}}^f(k, \theta_t)\|^{2}_{2}\right],
\label{eq:loss}
\end{equation}
where $\mathcal{B}$ represents the batch, and $g.t.$ denotes the groundtruth sinograms.
$\lambda=\|g.t. - \hat{\mathbf{C}}^f(k, \theta_t)\|$ is an adaptive regularization term to alleviate the overfitting brought by the ``coarse'' estimations occurred when the model is close to convergence.

\subsection{Multi-layer Perceptron Architecture} \label{sec:MLP}
\begin{figure}[!t]
  \centering
  \includegraphics[width=0.48\textwidth]{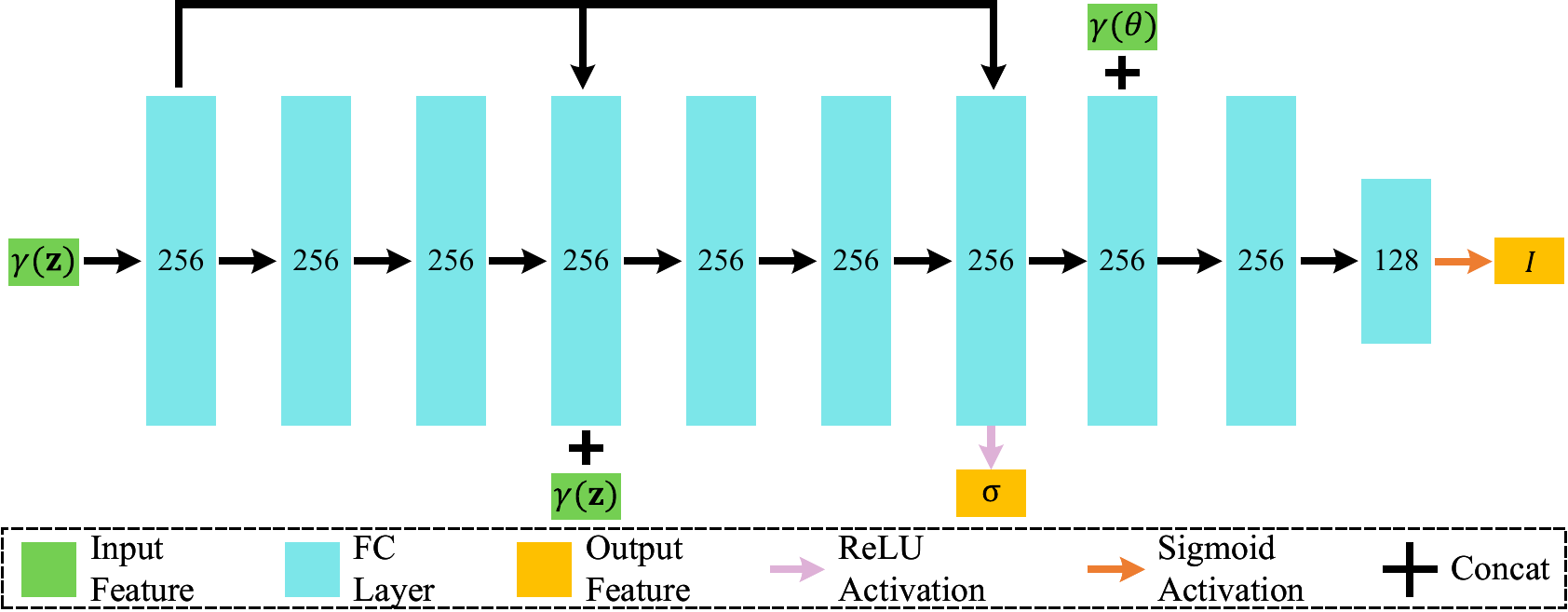}
  \caption{
    The architecture of our multi-layer perceptron (MLP) $F_{\Theta}$.
    For a given field coordinate $\mathbf{z}$ and projection angle $\theta$ encoded by the positional encoding $\gamma(\cdot)$ in \cref{eq:PE}, we first pass the coordinate vector $\gamma(\mathbf{z})$ through 9 fully-connected (FC) layers, each FC layer hasing $256$ channels.
    Then, to construct the residual connections, the feature of the 1st and 4th layers are added into the 4th and 7th layers, respectively.
    Since the attenuation coefficient $\sigma$ is only related to $\mathbf{z}$, $\sigma$ is predicted at the 7th layer with ReLU activations.
    Finally, we concatenate angle vector $\gamma(\theta)$ and downscale the feature channels into $128$, and the light intensity $I$ is predicted with Sigmoid activations.
  }
  \label{fig:MLP}
\end{figure}

To adapt the SVCT reconstruction, we reform the multi-layer perceptron (MLP) architecture of NeRF \cite{NeRF} as shown in \cref{fig:MLP}.
Visually, the input vector is split into the position coordinate $\mathbf{z}$ and projection angle $\theta$, and the output is a 2D union of $(I, \sigma)$.
The position coordinates are set to $\mathbf{z}$ is $(\hat{x}_{i}, \hat{y}_{i})$ and $(\hat{x}_{i}, \hat{y}_{i}, \hat{z}_{i})$ for 2D CT slices and 3D CT volumes, respectively.
Before feeding $(\mathbf{z}, \theta)$ into the MLP $F_{\Theta}$, each input vector $\omega$ is converted into higher dimension by the Fourier feature mapping $\gamma(\cdot, L)$ introduced in NeRF \cite{NeRF}:
\begin{equation}
\gamma(\omega, L)=\omega\bigcup^{L-1}_{i=0}(\sin(2^{i}\omega), \cos(2^{i}\omega)),\ where\ L\in\mathbb{N},
\label{eq:PE}
\end{equation}
where $L$ is a hyperparameter of frequency scale set to $10$ and $6$ for $\mathbf{z}$ and $\theta$, respectively.
As introduced in \cref{eq:volume}, $\sigma(\cdot)$ and $I(\mathbf{z}, \cdot)$ are the functions related to $\mathbf{z}$ and $(\mathbf{z}, \theta)$, respectively.
To decouple the features of $\sigma$ and $I$, we first feed $\mathbf{z}$ into the MLP $F_{\Theta}$ to predict the attenuation coefficient $\sigma$, and then concatenate the last layer and $\theta$ to predict the light intensity $I$.
Since the proposed rendering modules are differentiable, the dual MLPs $F^c_{\Theta}$ and $F^f_{\Theta}$ can be optimized by minimizing the rendering loss $\mathcal{L}$ introduced in \cref{eq:loss}.

\section{Experiments}\label{sec:exp}
In this section, we conduct extensive experiments and in-depth analysis to demonstrate our superiority.
The hyperparameters and model settings are consistent in all experiments.

\begin{table*}[!t]
  \setlength{\tabcolsep}{1.5mm}
  \caption{
  Comparisons on COVID-19 \cite{covid19} dataset for \bt{2D} SVCT reconstruction in terms of \bt{PSNR} / \bt{SSIM}. The re-projection technique for all the INR-based methods is FBP \cite{FBP}.
  \bt{Bold} and \ut{underline} texts indicate the best and second best performance.}
  \label{tab:2DSVCT}
  \centering
  \begin{tabular}{lccccccccc}
  \toprule
  \multirow{3}{*}{Methods}& \multicolumn{4}{c}{\xt{Parallel} X-ray beam} && \multicolumn{4}{c}{\xt{Fan} X-ray beam} \\\cmidrule{2-5}\cmidrule{7-10}
                          &15 \xt{views}           &30 \xt{views}           &45 \xt{views}           &60 \xt{views}           &&15 \xt{views}&30 \xt{views}&45 \xt{views}&60 \xt{views}\\\midrule
  \multicolumn{10}{l}{\scriptsize\textit{Conventional Methods}}\\
  FBP \cite{FBP}          &19.38      / 0.1761     &22.70      / 0.3039     &24.47      / 0.4043     &25.62      / 0.4794     &&20.29      / 0.2060     &23.98      / 0.3189     &26.18      / 0.4043     &26.67      / 0.4805     \\\midrule
  \multicolumn{10}{l}{\scriptsize\textit{CNN-based Methods}}\\          
  FBPConv \cite{FBPConv}  &28.76      / 0.6957     &31.58      / 0.7606     &32.18      / \ut{0.7810}&33.00      / \ut{0.8022}&&28.21      / 0.6958     &30.17      / 0.7012     &30.86      / 0.7250     &31.53      / 0.7680     \\
  FreeSeed \cite{freeseed}&\bt{29.91} / \ut{0.7085}&\ut{31.72} / \ut{0.7714}&\ut{32.25} / 0.7445     &\ut{33.71} / 0.7984     &&\ut{28.41} / \ut{0.6963}&\bt{31.03} / \ut{0.7563}&\bt{31.54} / \ut{0.7697}&\ut{31.98} / 0.7804     \\\midrule
  \multicolumn{10}{l}{\scriptsize\textit{INR-based Methods}}\\            
  GRFF \cite{GRFF}        &28.54      / 0.6813     &30.50      / 0.7268     &30.64      / 0.7321     &31.20      / 0.7563     &&28.00      / 0.6592     &29.31      / 0.6768     &30.08      / 0.6996     &30.53      / 0.7246     \\
  NeRP \cite{NeRP}        &29.32      / 0.6104     &31.00      / 0.7205     &31.48      / 0.7621     &32.15      / 0.7686     &&28.36      / 0.6463     &30.22      / 0.7227     &31.02      / 0.7505     &31.31      / 0.7690     \\
  CoIL \cite{CoIL}        &26.61      / 0.6176     &29.41      / 0.7061     &30.92      / 0.7556     &31.96      / 0.7863     &&25.99      / 0.6419     &28.40      / 0.7134     &30.79      / 0.7577     &31.06      / \ut{0.7860}\\
  SCOPE \cite{scope}      &23.99      / 0.5772     &27.38      / 0.6470     &29.87      / 0.7072     &31.58      / 0.7570     &&21.11      / 0.5278     &22.87      / 0.5595     &24.81      / 0.6385     &25.00      / 0.6577     \\
  \rowcolor{MyGray}
  \bt{CoCPF}  [Ours]      &\ut{29.37} / \bt{0.7130}&\bt{31.96} / \bt{0.7765}&\bt{33.24} / \bt{0.8029}&\bt{33.87} / \bt{0.8160}&&\bt{28.53} / \bt{0.6983}&\ut{30.85} / \bt{0.7573}&\ut{31.50} / \bt{0.7977}&\bt{32.04} / \bt{0.8202}\\\bottomrule
  \end{tabular}
\end{table*}

\begin{figure*}[!t]
  \centering
  \includegraphics[width=\textwidth]{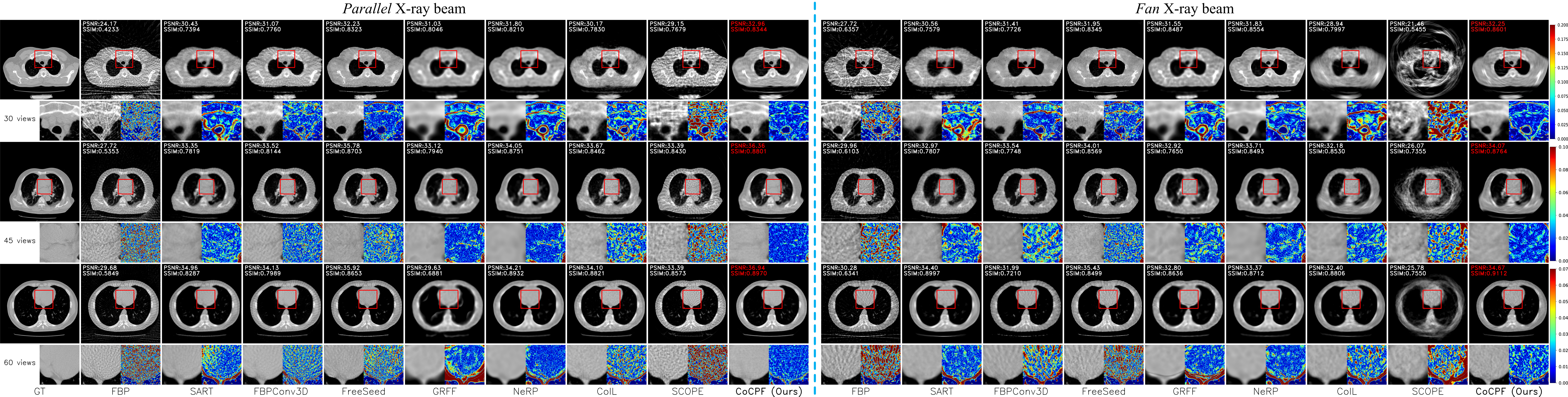}
  \caption{
    Comparisons against FBP \cite{FBP}, FBPConv \cite{FBPConv}, FreeSeed \cite{freeseed}, GRFF \cite{GRFF}, NeRP \cite{NeRP}, CoIL \cite{CoIL} and SCOPE \cite{scope} for \bt{2D} \xt{parallel}- and \xt{fan}-beam SVCT reconstructions on COVID-19 \cite{covid19} dataset.
    Subfigures (bottom left) show the anatomic structures zoomed in the boxes, and heatmaps (bottom right) display the difference related to the GT.
    \textcolor{red}{Red} text denotes the best score.
  }
  \label{fig:visual_2d}
  \end{figure*}

\subsection{Experimental Details}
\noindent\bt{Datasets.}
The experimental data are selected from 3 publicly available datasets, including 2 simulated datasets: COVID-19 \cite{covid19} and 2019 Kidney Tumor Segmentation Challenge (KiTS19) \cite{kits19}, and 1 real projection dataset: AAPM-LDCT-PD \cite{aapm}, which are discussed as follows:

\textbf{1)} \textbf{COVID-19} \cite{covid19} dataset contains CT scans from 1000+ subjects.
We randomly select 600 2D slices from 100 subjects, of which 500 ones for training CNN-based methods, 60 and the rest 40 for validation and test, respectively.
All the slices have the same image shape of $512\times512$.

\textbf{2)} \textbf{2019 Kidney Tumor Segmentation Challenge} (KiTS19) \cite{kits19} dataset consists of arterial phase abdominal CT scans from 300 unique kidney cancer subjects.
We randomly select 200 3D CT volumes, of which 150 ones as the training set for the CNN-based methods, 10 and the rest 40 as the validation and test sets, respectively.
Each CT volume is resized into $256\times256\times80$ image dimension.

\textbf{3)} \textbf{AAPM Low-Dose CT and Projection Data} (AAPM-LDCT-PD) \cite{aapm} dataset is a public CT dataset based on the 2016 Low Dose CT Grand Challenge, sponsored by the AAPM, NIBIB, and Mayo Clinic, released in The Cancer Imaging Archive (TCIA) \cite{TCIA}. This dataset comprises CT images and real projection data acquired by scanning the patients on the SOMATOM Definition Flash CT scanner from Siemens Healthcare.
All the sinograms are projected by the fan-beam geometry with the same detector size of $736$ related to a $512\times512$ CT image, and the geometry details are recorded in the metadata for re-projection.
For evaluating the INR-based methods, we randomly select 40 2D sinograms.

\noindent\bt{Dataset Simulation.}
\xt{For simulated data}, we follow the settings in \cite{NeRP} to generate the sinograms by projecting the raw CT images based on various projection geometries simulated by the Operator Discretization Library (ODL) \cite{odl} under different projection numbers (15, 30, 45, and 60).

\xt{For real projection data}, we uniformly select 60, 120, and 180 projection views from the full-view sinograms to generate the corresponding SV ones.
The groundtruth CT images are generated by applying FBP \cite{FBP} on the full-view sinograms with the geometries recorded in metadata, where the FBP \cite{FBP} is re-implemented by the TorchRadon library \cite{torch_radon}.

\xt{For fair comparisons}, all the INR-based methods are only trained on SV sinograms to reconstruct the CT images without any additional knowledge such as the longitudinal CT scans at the same positions of the same subjects used in \cite{NeRP}.

\noindent\bt{Implementation Details.} \label{sec:ID}
Our method is implemented on top of \cite{nerf-pytorch}, a Pytorch \cite{Pytorch} re-implementation of NeRF \cite{NeRF}.
Meanwhile, our experiments are also based on Pytorch \cite{Pytorch} framework, and run on a single GeForce RTX 3090 GPU with 24G memory.
\xt{CoCPF} first samples $N^c=64$ points for the coarse model $F^{c}_{\Theta}$ and feeds total $N^c+N^f=128$ points (the sorted union of $N^c=64$ coarse and $N^f=64$ fine samples) into the fine model $F^{f}_{\Theta}$.
The width $\varpi$ and length $\rho$ of the stripe $\mathcal{S}_{k}(\varpi, \rho, \theta_t)$ are set to $2$ and $\frac{2}{W}$, respectively.

\xt{For training}, we employ Adam \cite{adam} as the optimizer with a weight decay of $10^{-6}$ and a batch size of $2048$.
The maximum iteration is set to $20000$ and $40000$ for 2D slices and 3D volumes, respectively. 
The learning rate is annealed logarithmically from $2\times10^{-3}$ to $2\times10^{-5}$ during training.

\xt{For testing}, we first generate DV sinograms by uniformly sampling 720 partitions within $[0, \pi]$, and then apply FBP \cite{FBP} re-implemented by ODL \cite{odl} to reconstruct the CT images, where $N^c=N^f=8$ for speed-up.

\noindent\textbf{Runtime \& Parameters.}
The training time of \xt{CoCPF} for 2D SVCT reconstruction is nearly 10 minutes, while it takes about 25 minutes for 3D ones.
The inference time of a $512\times512$ CT scan is nearly 1 seconds, and it requires about 18 seconds for a $256\times256\times80$ CT volume.
The MLP is a lightweight model that only contains 0.53M parameters.

\noindent\bt{Evaluation Metrics.}
Following the experimental settings in \cite{CoIL}, we use PSNR and SSIM \cite{ssim} re-implemented by scikit-image \cite{skimage} packages to evaluate the performance.

\begin{figure*}[!t]
  \centering
  \includegraphics[width=\textwidth]{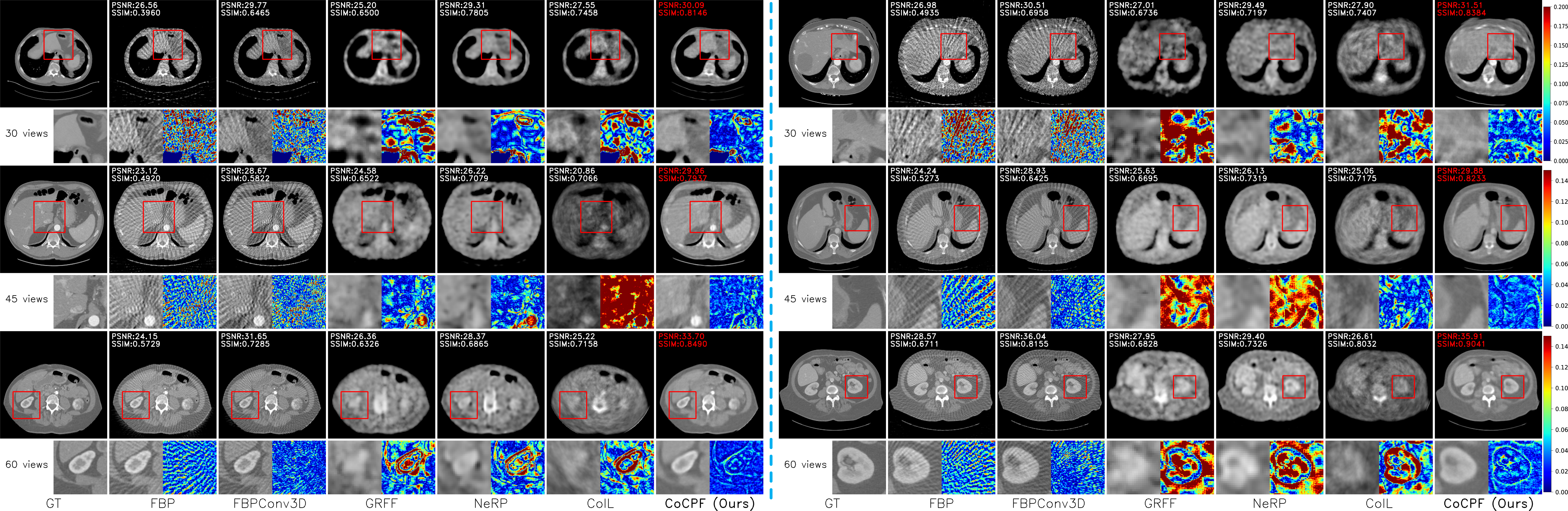}
  \caption{
    Comparisons against FBP \cite{FBP}, FBPConv3D \cite{FBPConv}, GRFF \cite{GRFF}, NeRP \cite{NeRP} and CoIL \cite{CoIL} for \bt{3D} \xt{parallel}-beam SVCT reconstruction on KiTS19 \cite{kits19} dataset within $[-500HU, 500HU]$.
    Subfigures (bottom left) show the anatomic structures zoomed in the boxes, and heatmaps (bottom right) display the difference related to the GT.
    \textcolor{red}{Red} text denotes the best score.
  }
  \label{fig:visual_3d}
\end{figure*}

\subsection{Comparison with State-of-the-art Methods}\label{sec:comp}
We compare the proposed \xt{CoCPF} with 7 state-of-the-art methods, including 1 conventional methods: FBP \cite{FBP}, 2 fully-supervised CNN-based methods: FBPConv \cite{FBPConv} and FreeSeed \cite{freeseed}, and 4 recent INR-based methods: CoIL \cite{CoIL}, GRFF \cite{GRFF}, NeRP \cite{NeRP} and SCOPE \cite{scope}.
Specifically, FBP is re-implemented by ODL \cite{odl}, and we follow its setting for the re-projections.
For fully-supervised methods: FBPConv \cite{FBPConv} and FreeSeed \cite{freeseed} are trained on COVID-19 \cite{covid19} dataset for 2D SVCT reconstructions, while FBPConv3D re-implemented by 3D U-Net is trained on KiTS19 \cite{kits19} dataset for 3D ones.
For the INR-based methods, the results of CoIL \cite{CoIL}, SCOPE \cite{scope}, GRFF \cite{GRFF}, NeRP \cite{NeRP} and our \xt{CoCPF} are generated by their official implementations on the SV sinograms without any additional knowledge and post-processings.

\begin{table}[!t]
  \setlength{\tabcolsep}{1.3mm}
  \caption{
    Comparisons on KiTS19 \cite{kits19} dataset for \bt{3D} \xt{parallel}-beam SVCT reconstruction in terms of \bt{PSNR} / \bt{SSIM}. The re-projection technique is FBP \cite{FBP}.
  \bt{Bold} and \ut{underline} texts indicate the best and second best performance.}
  \scriptsize
  \label{tab:3DSVCT}
  \centering
  \begin{tabular}{lcccc}
  \toprule
  Methods&15 \xt{views}           &30 \xt{views}           &45 \xt{views}           &60 \xt{views}           \\\midrule
  \multicolumn{5}{l}{\textit{Conventional Methods}}\\
  FBP \cite{FBP}          &20.71      / 0.3683     &23.46      / 0.5389     &24.74      / 0.6324     &25.49      / 0.6923\\\midrule
  \multicolumn{5}{l}{\textit{CNN-based Methods}}\\           
  FBPConv3D \cite{FBPConv}&28.32      / 0.6937     &\ut{31.46} / 0.7599     &\ut{31.90} / 0.7786     &\ut{32.53} / 0.8202     \\\midrule
  \multicolumn{5}{l}{\textit{INR-based Methods}}\\           
  GRFF \cite{GRFF}        &27.72      / 0.7379     &29.56      / \ut{0.8046}&29.83      / \ut{0.8455}&29.74      / 0.8300     \\
  NeRP \cite{NeRP}        &\ut{29.00} / \ut{0.7502}&29.44      / 0.7915     &30.13      / 0.8436     &30.89      / \ut{0.8587}\\
  CoIL \cite{CoIL}        &27.11      / 0.6702     &28.23      / 0.7184     &28.86      / 0.7466     &29.15      / 0.7573     \\
  \rowcolor{MyGray}
  \bt{CoCPF} [Ours]       &\bt{29.88} / \bt{0.7989}&\bt{32.50} / \bt{0.8720}&\bt{33.53} / \bt{0.8947}&\bt{33.85} / \bt{0.9012}\\\bottomrule
  \end{tabular}
\end{table}

\noindent\bt{Quantitative Comparison.}
We report the performance for 2D and 3D SVCT reconstructions under the \xt{parallel} and \xt{fan} X-ray beams on the simulated datasets COVID-19 \cite{covid19} and KiTS19 \cite{kits19} in \Cref{tab:2DSVCT,tab:3DSVCT}, respectively.
As shown, the proposed \xt{CoCPF} favorably surpasses the state-of-the-art INR- and CNN-based methods, achieving the consistently preferable performance for 2D and 3D SVCT reconstructions under various projection numbers and geometries. To investigate our performance on the real projection data, we further it against state-of-the-art INRs on AAPM-LDCT-PD \cite{aapm} in \Cref{tab:AAPM}
As reported, the INR-based methods only achieve unsatisfactory reconstruction results, while our \xt{CoCPF} acquire high-quality CT images under the various projection numbers, suggesting the correctness of our motivation discussed in \Cref{sec:real}.
\textit{Moreover}, \textit{CoCPF} only needs to train on a single SV sinogram for a short time, which can be applied to practical scenarios.

\begin{table}[!t]
  \setlength{\tabcolsep}{1.25mm}
  \caption{
  Comparisons on AAPM-LDCT-PD \cite{aapm} for \bt{2D} \xt{fan}-beam SVCT reconstruction.
  \bt{Bold} text indicates the best performance. The re-projection technique is FBP \cite{FBP}.}
  \label{tab:AAPM}
  \centering
  \begin{tabular}{lcccccccc}
  \toprule
  \multirow{3}{*}{Methods} & \multicolumn{2}{c}{60 \textit{views}} & & \multicolumn{2}{c}{120 \textit{views}} & & \multicolumn{2}{c}{180 \textit{views}}\\\cmidrule{2-3}\cmidrule{5-6}\cmidrule{8-9}
                       &PSNR\ua   &SSIM\ua    &&PSNR\ua   &SSIM\ua    &&PSNR\ua   &SSIM\ua    \\\midrule
\multicolumn{8}{l}{\scriptsize\textit{Conventional Methods}}\\
FBP \cite{FBP}           &16.38     &0.4186     &&18.11     &0.5015     &&19.80     &0.5703     \\\midrule
\multicolumn{8}{l}{\scriptsize\textit{INR-based Methods}}\\
GRFF \cite{GRFF}         &19.90     &0.6076     &&21.73     &0.6628     &&23.59     &0.6983     \\
NeRP \cite{NeRP}         &21.48     &0.6546     &&22.42     &0.6934     &&24.71     &0.7354     \\
CoIL \cite{CoIL}         &23.09     &0.7165     &&25.83     &0.7864     &&27.31     &0.8786     \\
SCOPE \cite{scope}       &18.20     &0.5116     &&18.33     &0.5270     &&19.04     &0.5467     \\
\rowcolor{MyGray}
\bt{CoCPF} [Ours]         &\bt{28.18}&\bt{0.8572}&&\bt{30.97}&\bt{0.8818}&&\bt{33.28}&\bt{0.9042}\\\bottomrule
  \end{tabular}
\end{table}

\begin{table*}[!t]
  \caption{Comparisons of DV sinogram synthesis under various projection numbers in terms of PSNR on \bt{2D} SVCT datasets: COVID-19 \cite{covid19} and AAPM-LDCT-PD \cite{aapm}.
  \bt{Bold} text indicates the best performance.}
  \label{tab:DVSR}
  \setlength{\tabcolsep}{1.7mm}
  \centering
  \begin{tabular}{lccccccccccccc}
  \toprule
  \multirow{4}{*}{Methods} & \multicolumn{9}{c}{COVID-19 \cite{covid19}} && \multicolumn{3}{c}{AAPM-LDCT-PD \cite{aapm}}\\\cmidrule{2-10}\cmidrule{12-14}
  & \multicolumn{4}{c}{\xt{Parallel} X-ray beam} && \multicolumn{4}{c}{\xt{Fan} X-ray beam} && \multicolumn{3}{c}{\xt{Fan} X-ray beam}\\\cmidrule{2-5}\cmidrule{7-10}\cmidrule{12-14}
  & 15 \xt{views} & 30 \xt{views} & 45 \xt{views} & 60 \xt{views} && 15 \xt{views} & 30 \xt{views} & 45 \xt{views} & 60 \xt{views} && 60 \xt{views} & 120 \xt{views} & 180 \xt{views} \\\midrule
  Bicubic            &31.60     &38.65    &42.75     &44.59      &&27.10     &36.48     &41.34     &43.59     &&31.87     &33.98     &36.64      \\
  CoIL \cite{CoIL}   &36.47     &41.98    &44.83     &47.05      &&34.24     &42.50     &44.89     &46.05     &&33.92     &35.18     &39.19      \\
  SCOPE \cite{scope} &24.14     &40.03    &44.22     &46.49      &&--        &--        &--        &--        &&21.29     &24.79     &25.17      \\
  \rowcolor{MyGray}
  \bt{CoCPF} [Ours]  &\bt{41.84}&\bt{48.36}&\bt{50.26}&\bt{51.75}&&\bt{41.73}&\bt{46.24}&\bt{49.16}&\bt{51.79}&&\bt{38.27}&\bt{45.08}&\bt{50.41}\\\bottomrule
  \end{tabular}
\end{table*}

\begin{figure*}[!t]
  \centering
  \includegraphics[width=\textwidth]{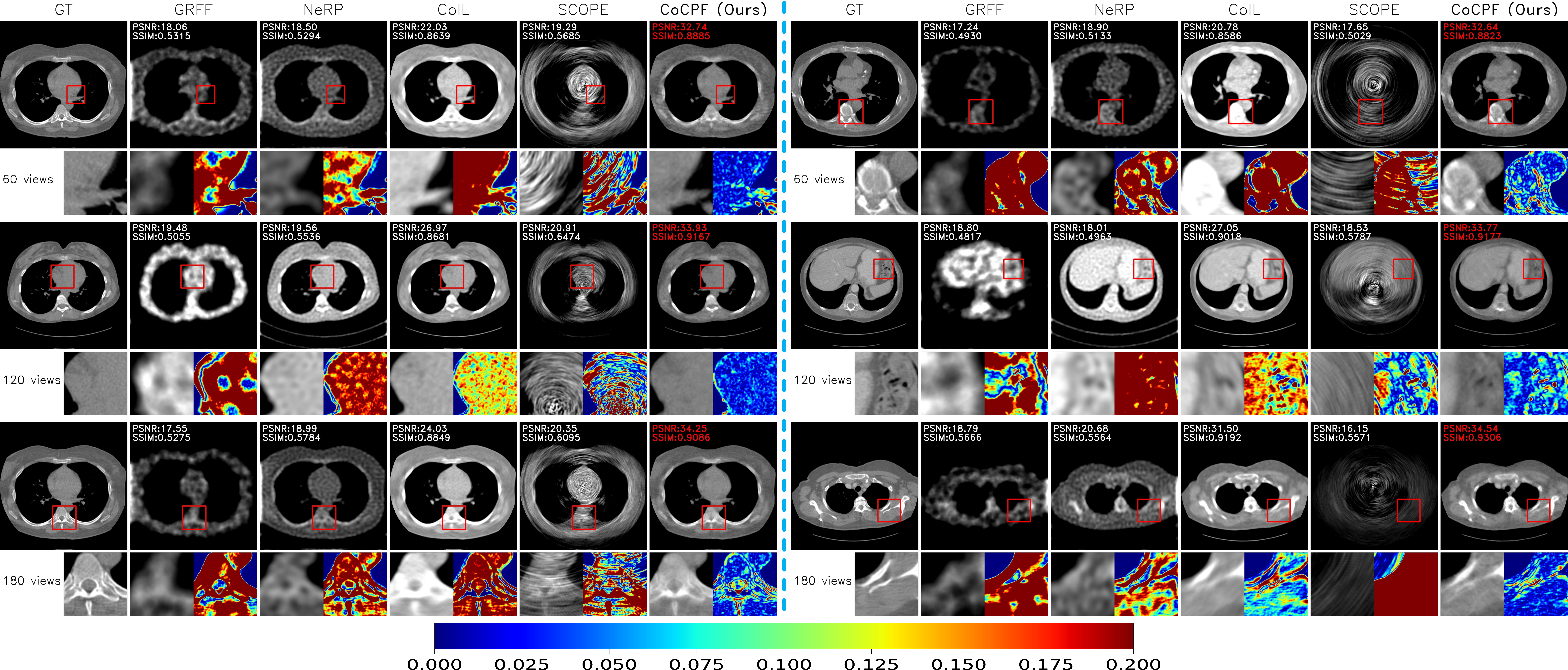}
  \caption{
    Comparisons against GRFF \cite{GRFF}, NeRP \cite{NeRP}, CoIL \cite{CoIL} and SCOPE \cite{scope} for \bt{2D} \textit{fan}-beam SVCT reconstruction on AAPM-LDCT-PD \cite{aapm} dataset within $[-500HU, 500HU]$.
    Subfigures (bottom left) show the anatomic structures zoomed in the boxes, and heatmaps (bottom right) display the difference related to the GT.
    \textcolor{red}{Red} text denotes the highest score.
  }
  \label{fig:visual_aapm}
\end{figure*}

\noindent\bt{Visual Comparison.}
We visualize the reconstruction results of \xt{CoCPF} and other competitors for 2D and 3D SVCT reconstruction in \cref{fig:visual_2d,fig:visual_3d} under various projection numbers and geometries, respectively.
As shown, GRFF \cite{GRFF} and NeRP \cite{NeRP} struggle to accurately reconstruct the anatomic structures, while CoIL \cite{CoIL} and SCOPE \cite{scope} preserve more details but also produce severe artifacts.
We further visualize the results between existing INR-based methods and our \textit{CoCPF} on real projection data in \cref{fig:visual_aapm}.
Visually, GRFF \cite{GRFF}, NeRP \cite{NeRP}, and SCOPE \cite{scope} struggle to reconstruct high-quality CT images from the sub-sampled projection data, while CoIL \cite{CoIL} cannot generate accurate details, producing blurry contents in their results. 
These unsatisfactory results prove that the holes formed in their representation fields affect the performance of exhibited INR-based methods, leading to sub-optimal results.
Compared with the exhibited methods, \xt{CoCPF} favorably surpasses the above INR-based methods and achieves comparable image quality to the fully-supervised CNN-based methods: FBPConv \cite{FBPConv} and FreeSeed \cite{freeseed}, yielding better visual verisimilitude to the GT images with fine-grained details and fewer artifacts.
These visualization results are consistent with the performance reported in \Cref{tab:2DSVCT,tab:3DSVCT,tab:AAPM}, proving the correctness of our motivation.

\noindent\textbf{Comparisons on DV Sinogram Synthesis.}
To directly validate the reliability of the DV sinograms synthesized by our \textit{CoCPF}, \textit{i.e.,} whether the synthesized dense-view sinograms match the real full-view ones or not, we evaluate their difference on COVID-19 \cite{covid19} and KiTS19 \cite{kits19} datasets in terms of PSNR in \Cref{tab:DVSR}.
As shown, our \textit{CoCPF} achieves the most similarity to the full-view projections under various projection numbers, suggesting that the DV sinograms estimated by our \textit{CoCPF} are more approximated to the groundtruths, thereby \textit{CoCPF} can achieve superior reconstruction quality to the competitors after re-projection.
It also demonstrates that the holes are well-filled by the proposed modules, proving the correctness of our motivation.

\begin{table*}[!t]
  \setlength{\tabcolsep}{1.65mm}
  \caption{Comparisons against various ablations for \bt{2D} and \bt{3D} \xt{parallel}-beam SVCT reconstructions on COVID-19 \cite{covid19} and KiTS19 \cite{kits19} datasets, respectively.
  \bt{Bold} text indicates the best performance in this table.
  }
  \label{tab:ab}
  \centering
  \footnotesize
  \begin{tabular}{ccccccccccccccc}
  \toprule
  \multirow{4}{*}{\xt{SVS}} & \multirow{4}{*}{\xt{PVR}} & \multirow{4}{*}{\xt{SHR}} & \multirow{4}{*}{$\mathcal{L}$} & \multicolumn{5}{c}{COVID-19 \cite{covid19}} && \multicolumn{5}{c}{KiTS19 \cite{kits19}} \\\cmidrule{5-9}\cmidrule{11-15}
      &   &   &   &\multicolumn{2}{c}{30 \xt{views}}      &&\multicolumn{2}{c}{60 \xt{views}}&&\multicolumn{2}{c}{30 \xt{views}}&&\multicolumn{2}{c}{60 \xt{views}} \\\cmidrule{5-6}\cmidrule{8-9}\cmidrule{11-12}\cmidrule{14-15}
      &   &   &   &PSNR\ua             &SSIM\ua                &&PSNR\ua             &SSIM\ua                &&PSNR\ua             &SSIM\ua                &&PSNR\ua             &SSIM\ua\\\midrule
      &   &   &   &28.85               &0.7020                 &&31.73               &0.7610                 &&30.14               &0.8415                 &&32.11               &0.8595 \\
  \ck&   &   &   &30.08+\df{1.23}     &0.7344+\df{0.0324}     &&32.64+\df{0.91}     &0.7834+\df{0.0224}     &&31.38+\df{1.24}     &0.8534+\df{0.0119}     &&32.87+\df{0.76}     &0.8814+\df{0.0219} \\
  \ck&\ck&   &   &31.25+\df{1.17}     &0.7642+\df{0.0298}     &&33.33+\df{0.69}     &0.8007+\df{0.0173}     &&31.91+\df{0.53}     &0.8683+\df{0.0149}     &&33.54+\df{0.67}     &0.8957+\df{0.0143} \\
  \ck&\ck&\ck&   &31.61+\df{0.36}     &0.7741+\df{0.0099}     &&33.72+\df{0.39}     &0.8105+\df{0.0098}     &&32.25+\df{0.34}     &0.8707+\df{0.0024}     &&33.80+\df{0.26}     &0.9007+\df{0.0050} \\
  \rowcolor{MyGray}
  \ck&\ck&\ck&\ck&\bt{31.96}+\df{0.35}&\bt{0.7765}+\df{0.0024}&&\bt{33.87}+\df{0.15}&\bt{0.8160}+\df{0.0055}&&\bt{32.50}+\df{0.25}&\bt{0.8720}+\df{0.0013}&&\bt{33.85}+\df{0.05}&\bt{0.9012}+\df{0.0005} \\\hline
  \end{tabular}
\end{table*}

\subsection{Ablation Study \& Parametric Sensitivity Analysis} \label{sec:ab}
In this subsection, we conduct comprehensive experiments to prove the correctness of \xt{CoCPF}'s design.
We first carry out an ablation study to investigate the effectiveness of the proposed modules.
Subsequently, we evaluate the \xt{CoCPF}'s parametric sensitivity under different parameter settings.

\noindent\bt{\xt{CoCPF}'s Ablation Variants.}
We evaluate against the ablations of \xt{CoCPF} with each module, including \xt{stripe-based volume sampling}, \xt{piecewise-consistent volume rendering}, \xt{stripe-based hierarchical rendering}, and the rendering loss \cref{eq:loss}, which are abbreviated as \xt{SVS}, \xt{PVR}, \xt{SHR} and $\mathcal{L}$ in \Cref{tab:ab}, respectively.
The baseline here is the model that employs ray sampling and summation operation, where the $\lambda$ in of rendering loss is set to 1.
For fair comparisons, each ablation variant is trained with the same experimental settings, including the same architecture of the MLP $F_{\Theta}$ introduced in \Cref{sec:MLP}.
As shown, the performance of the baseline model (row 1) can be significantly improved by employing \xt{SVS} (row 2), suggesting sampling stripes instead of rays can address the hole-forming issues.
Compared to the summation formula, employing the proposed \xt{PVR} can obtain better reconstruction quality (row 3).
Since the one-stage result is a coarse estimation, the proposed \xt{SHR} (row 4) and $\mathcal{L}$ (row 5) can further improve the performance and alleviate the distraction brought by the coarse terms, respectively.

\noindent\bt{\xt{CoCPF} under Different Parameteric Settings.}
We evaluate the performance of \xt{CoCPF} under different parameteric settings: ``$\varpi=\frac{1}{W}$'' and ``$\varpi=\frac{4}{W}$'' represent to the with of stripe.
The default setting of \xt{CoCPF} is introduced in \Cref{sec:ID}, where $\varpi=\frac{2}{W}$.
As reported in \Cref{tab:ps}, the default setting achieves consistent outperformance at various projection numbers, and ``$\varpi=\frac{1}{W}$'' and ``$\varpi=\frac{4}{W}$'' are more suitable for 30 and 60 projection views, respectively.
Since the performance of different settings is comparable to each other, it suggests \xt{CoCPF} is a parameter-insensitive method with good robustness against different parametric settings.

\subsection{Analysis}
We compare our \textit{CoCPF} with 7 state-of-the-art methods in \Cref{sec:comp}.
The experimental results of quantitative and visual comparisons on simulated and real projection data demonstrate that \textit{CoCPF} achieves consistently superior performance to the competitors under various projection numbers and geometries for 2D and 3D SVCT reconstructions, which confirms the correctness of our motivation and model designs.
Moreover, compared with existing \textit{view-synthesis} INRs, the DV sinograms synthesized by our \xt{CoCPF} are much closer to the groundtruths, thereby \xt{CoCPF} can acquire higher-quality CT images than the competitors after re-projection.
It also indicates that the holes are well-filled during optimization, which means \xt{CoCPF} successfully addresses the hole-forming issues, proving the correctness of our motivation.
Furthermore, we also thoroughly evaluate the performance of our \textit{CoCPF} under different ablation variants, and parametric settings in \Cref{sec:ab}.
The ablation results demonstrate the effectiveness of our model designs, while the consistent preferable performance under different parametric settings suggests our \textit{CoCPF} is a parameter-insensitive method.
Compared to state-of-the-art methods, \textit{CoCPF} achieves superior performance and better robustness against various situations.

\begin{table}[!t]
  \setlength{\tabcolsep}{0.5mm}
  \caption{
  Comparisons under different parametric settings for \bt{2D} and \bt{3D} \xt{parallel}-beam SVCT reconstructions on COVID-19 \cite{covid19} and KiTS19 \cite{kits19} datasets, respectively.
  \bt{Bold} text indicates the best performance in this table.
  }
  \label{tab:ps}
  \centering
  \footnotesize
  \begin{tabular}{cccccccccccc}
  \toprule
  \multirow{4}{*}{Settings} & \multicolumn{5}{c}{COVID-19 \cite{covid19}} && \multicolumn{5}{c}{KiTS19 \cite{kits19}}  \\\cmidrule{2-6}\cmidrule{8-12}
                        &\multicolumn{2}{c}{30\xt{views}}&&\multicolumn{2}{c}{60 \xt{views}}&&\multicolumn{2}{c}{30 \xt{views}}&&\multicolumn{2}{c}{60 \xt{views}} \\\cmidrule{2-3}\cmidrule{5-6}\cmidrule{8-9}\cmidrule{11-12}
                        &PSNR\ua   &SSIM\ua              &&PSNR\ua   &SSIM\ua               &&PSNR\ua   &SSIM\ua               &&PSNR\ua   &SSIM\ua     \\\midrule
    \rowcolor{MyGray}
    default             &31.96     &0.7765               &&33.87     &\bt{0.8160}           &&\bt{32.50}&0.8720                &&33.85     &0.9012      \\
    $\varpi=\frac{1}{W}$&31.90     &0.7742               &&\bt{33.90}&0.8152                &&32.13     &0.8714                &&\bt{33.94}&\bt{0.9027} \\
    $\varpi=\frac{4}{W}$&\bt{32.00}&\bt{0.7771}          &&33.75     &0.8154                &&32.46     &\bt{0.8729}           &&33.64     &0.8954      \\\bottomrule
  \end{tabular}
\end{table}

\section{Conclusion}\label{sec:con}
In this paper, we observed that existing INR-based methods suffer from hole-forming issues due to the ill-posedness brought by the sparsely-sampled measurements, leading to sub-optimal results.
Based on our findings, we propose a novel self-supervised method for SVCT reconstruction -- \bt{Co}ordinate-based \bt{C}ontinuous \bt{P}rojection \bt{F}ield (CoCPF), which can build a hole-free representation field for better reconstruction quality.
Specifically, to fill the holes, \xt{CoCPF} broadens the sampling regions from rays (1D space) to stripes (2D space) using the proposed \xt{stripe-based volume sampling} module.
Since the internal regions between adjacent SV projections can be well-covered, the holes are jointly optimized during training, reducing the ill-posed levels.
By feeding the sampling regions into the proposed differentiable rendering modules, \xt{CoCPF} can synthesize accurate DV sinograms and acquire high-quality CT images after re-projection.
Comprehensive experiments on simulated and real projection data for 2D and 3D SVCT reconstructions demonstrate that \xt{CoCPF} surpasses state-of-the-art INR- and CNN-based methods for 2D and 3D SVCT reconstruction, yielding better visual verisimilitude and fewer artifacts under various projection numbers and geometries.

\bibliographystyle{IEEEtran}
\bibliography{main}
\end{document}